# Femtosecond structural transformation of phase-change materials far from equilibrium monitored by coherent phonons

Muneaki Hase[1], Paul Fons[2], Kirill Mitrofanov[2], Alexander V. Kolobov[2] & Junji Tominaga[2]

[1]*Division of Applied Physics, Faculty of Pure and Applied Sciences, University of Tsukuba, 1-1-1 Tennodai, Tsukuba 305-8573, Japan.*

[2]*Nanoelectronics Research Institute, National Institute of Advanced Industrial Science and Technology, Tsukuba Central 4, 1-1-1 Higashi, Tsukuba 305-8562, Japan.*

Correspondence and requests for materials should be addressed to M. H. (email: mhase@bk.tsukuba.ac.jp)

**Multi-component chalcogenides, such as quasi-binary GeTe-$Sb_2Te_3$ alloys, are widely used in optical data storage media in the form of rewritable optical discs. $Ge_2Sb_2Te_5$ (GST) in particular has proven to be one of the best-performing materials, whose reliability allows more than $10^6$ write-erase cycles. Despite these industrial applications, the fundamental kinetics of rapid phase-change in GST remain controversial and active debate continues over the ultimate speed limit. Here we explore ultrafast structural transformation in a photo-excited GST superlattice, where GeTe and $Sb_2Te_3$ are spatially separated, using coherent phonon spectroscopy with pump-pump-probe sequences. By analysing the coherent phonon spectra in different time regions, complex structural dynamics upon excitation are observed in GST superlattice (but not in GST alloys), which can be described as the mixing of Ge sites from two different coordination environments. Our results suggest possible applicability of GST superlattice for ultrafast switching devices.**

The rapid phase change in GST materials involved in the writing (from crystalline to amorphous) and erasing (from amorphous to crystalline) of data in optical recording media is induced by irradiation with focused nanosecond laser pulses, leading to melt-quenching (amorphization) and annealing (crystallization), respectively[1-3]. The same concept has been applied to electrical memory with electrical pulses replacing optical pulses and resistance differences replacing reflectivity changes[4]. Thus, the conventional understanding of the dynamics of rapid phase change in optical recording media using GST materials is based upon a thermal process, limiting the speed of the write-erase cycle to the nanosecond range[1,2]. In electronic memories, the terms SET and RESET phases are used to describe the two structural phases[4,5]. By analogy, we use the same terminology here, although it is not commonly used with respect to optical memory. SET refers to the local structure in the crystalline phase of GST, typically characterized by resonantly bonded atoms with three shorter and three longer bonds, also referred to as octahedrally coordinated[6,7] when differences in the bond lengths are neglected, and RESET refers to the covalently bonded amorphous phase, described as having tetrahedral[8], defective octahedral[9], or pyramidal[10] Ge sites. The microscopic mechanism of rapid phase change in GST has been examined using optical absorption[11], Raman scattering, and x-ray absorption fine structure (XAFS) measurements[8], as well as density functional theory (DFT)-based *ab-initio* molecular dynamics (MD) simulations[3,12,13].

Interfacial phase change memory (iPCM) consists of a superlattice (SL) structure formed from alternating layers of GeTe and $Sb_2Te_3$ (refs. 7,14). iPCM was designed to utilize a solid-solid phase transformation between the covalently bonded (RESET) and resonantly bonded (SET) phases, induced predominantly by the displacement of Ge atoms at the interface, to achieve both faster and lower power threshold switching than in conventional GST alloys[7]. iPCM structures with certain thicknesses and atomic order in the individual blocks were also argued to be topological insulators, and recently, it was reported that iPCM could be switched between the Dirac-semimetal and gapped phases[15], an effect that has the potential to lead to novel spin memory devices and also as a platform to study the topological properties of superlattices.

Non-thermal phase transitions (electronic excitation induced phase transitions[16,17]) induced via strong photo-excitation by femtosecond laser pulses with sub-picosecond time resolution have been reported for a wide variety of materials including semiconductors[16], manganites[18] and insulators[19]. Recently, the possibility of non-thermal phase change has been theoretically[13,17] and experimentally[20,21] proposed for GST alloys. In an important step, Fons *et al.* reported based upon time-resolved XAFS that non-thermal optically-induced contributions to the amorphization in GST alloy may be present on sub-nanosecond time scales[21]. As a further step, using weak femtosecond laser pulses with pump fluences below 100 µJ cm$^{-2}$, Makino *et al.* demonstrated that for a prototypical iPCM structure, $[Ge_2Te_2/Sb_2Te_3]_{20}$, a phase change from the RESET into the SET phase could be induced by selectively exciting a phonon mode that involves Ge atoms using a double-pulse excitation[22]. Despite recent activity investigating non-thermal phase transitions in GST alloys and iPCM, the mechanism of the non-thermal phase transformation remains largely unknown, especially at sub-picosecond time scales under strong photo-excitation.

In the following, we report on systematic studies of the structural dynamics in a prototypical iPCM structure, [$Ge_2Te_2/Sb_2Te_3$]$_{20}$, using coherent phonon spectroscopy under strong photo-excitation employing both single and double pump-pulse excitation. In the SET phase of iPCM, we observe dramatic differences between the pre-transitional dynamics induced by single and double-pulse excitation. Under single pulse excitation, only phonon softening is observed for an optical phonon mode involving Ge-Te bonds. In contrast, under double-pulse excitation the optical phonon frequency exhibits an unexpected blue-shift (hardening), accompanied by the appearance of an additional peak in the coherent phonon spectra when the interval between the two pump-pulses is shorter than a few picoseconds and exceeds a critical total pump fluence. Furthermore, we find that the structural change dynamics in the photo-excited SET phase of iPCM are reversible, while those in the RESET phase of iPCM are irreversible. A metastable pre-phase transformation state with characteristic coherent phonon spectra in the SET phase of iPCM at ultrafast time scales suggests the presence of two different coordination environments around Ge atoms. Exploration of the non-thermal pre-phase transformations in GST materials will enable a deeper understanding of the local lattice structure far from equilibrium, and will potentially make it possible to increase the speed limit of switching in the phase change process beyond the current limit of nanoseconds down to sub-picosecond time scales.

**Results**

**Transient reflectivity studies of the SET phase of iPCM.** Figure 1a depicts a schematic view of the pump–pump–probe experiment[22,23]. To explore the coherent phonon spectra under non-equilibrium conditions prepared by photo-excitation (Fig. 1b), a stronger initial prepump-pulse ($P_1$ = 10.6 mJ cm$^{-2}$) promotes the sample into the excited state, followed by the generation of a coherent phonon in the excited state by another weaker pump ($P_2$ = 6.9 mJ cm$^{-2}$), which is monitored by a probe pulse ($P_3$ = 0.2 mJ cm$^{-2}$). The weaker pump ($P_2$) acts as a control pulse to coherently launch the vibrational amplitude beyond the threshold of transient local phase change around the Ge atoms[22]; use of a $P_2$ pulse fluence less than 2.0 mJ cm$^{-2}$ does not induce phonon hardening (Supplementary Fig. 1). The choice of fluences $P_1$ = 10.6 mJ cm$^{-2}$ and $P_2$ = 6.9 mJ cm$^{-2}$ was made because of the low-limit threshold for the observation of the phonon hardening in iPCM as discussed in the later and also to allow for matching the phonon amplitude resulting from the initial excitation at the time delay $\tau$ = 0 fs with the phonon amplitude induced by the second excitation pulse at the separation time ($\Delta t$) of $\Delta t$ = 290 – 870 fs (see Methods). The time delay ($\tau$) between the second pump and the probe pulse was scanned over the full range of the signal, for a variety of separation times ($\Delta t$) between $P_1$ and $P_2$ pulses.

Figure 2a shows the transient reflectivity ($\Delta R/R$) observed in the SET phase of an iPCM film without prepump-pulse ($P_1$) excitation (see the bottom trace) and five major traces observed for $P_1$ excitations at different $\Delta t$. One can notice that the coherent phonon oscillations in the excited state, highlighted by the rectangular light-blue region, change dramatically as the value of $\Delta t$ is varied; in particular, at $\Delta t$ = 290 – 870 fs, the coherent phonon after the arrival of the $P_2$ pulse exhibits a strongly damped oscillation. To compare the $\Delta R/R$ signal observed in iPCM under irradiation with double pump-pulses with that occurring for irradiation with a single pump-pulse, Figure 2b presents the transient reflectivity detected in the SET phase of an iPCM film by irradiation with a single $P_2$ pulse at different pump fluences. The coherent phonon oscillations exhibited an increase in amplitude with moderate damping of the phonon oscillation when the $P_2$ pulse fluence was increased up to 18.6 mJ cm$^{-2}$. To explore the phonon dynamics in the frequency domain, coherent oscillations occurring only after the arrival of the $P_2$ pulse, *i.e.*, the time region highlighted by the rectangular light-blue region in Fig. 2a (the entire region for the case of a single $P_2$ pulse in Fig. 2b), were Fourier transformed (FT)[24].

Without the $P_1$ excitation, the frequency of the coherent phonon was $\Omega_{SET}$ = 3.48 THz (see Fig. 3a), and was slightly red-shifted from the literature value[22,25], but remained consistent with the optical mode due to the local Ge-Te bonds of the resonantly bonded structure[22,25]. Similar phonon softening induced by single pump excitation has also been observed in conventional GST alloy films[26]. The choice of the value of $\Delta t$ is based on the period of the coherent phonon observed without prepump-pulse excitation; 3.48 THz corresponds to ≈ 290 fs and therefore multiples of 290 fs were used to excite the SET phase to coherently drive the vibrational amplitude beyond the threshold of transient structural change[22,27,28,29] by the $P_2$ pulse. As shown in Fig. 3b, at $\Delta t$ = 290 fs the original peak at 3.48 THz unexpectedly splits into two peaks, one 'blue-shifts' to 3.70 THz and an additional peak at 2.55 THz (labelled by ⋆) appears. The lower-frequency peak position shifts to higher frequency with increasing separation time $\Delta t$ and gradually disappears after $\Delta t$ = 1,740 fs. By contrast, the peak at 3.70 THz observed in the excited state is very close to the frequency of the RESET phase ($\Omega_{RESET}$ = 3.74 THz; see Supplementary Fig. 2) and gradually red-shifts to 3.57 THz at $\Delta t$ = 1,740 fs. After the aforementioned pump–pump–probe experiment for $\Delta t$ = 6,090 fs, the coherent phonon spectra was taken at the same sample position without the prepump-pulse ($P_1$), confirming that the excited state recovers to the original SET phase as the peak frequency is observed at the original frequency of 3.48 THz (Fig. 3c). Thus, the structural change observed in the photo-excited SET phase is reversible, but the excited crystal lattice exhibits a characteristic double-peak FT spectra on ultrafast time scales from $\Delta t$ = 290 to 1,160 fs.

For the case of single pulse excitation, different pre-transitional dynamics are observed as shown in Fig. 3d. The FT spectra obtained from the time-domain data in Fig. 2b exhibit a peak frequency for the optical mode due to the Ge-Te bonds at 3.52 THz for the single pump fluence of 5.3 mJ cm$^{-2}$, while at 18.6 mJ cm$^{-2}$ it red-shifts to 3.27 THz with a concomitant broadening of the peak structure. However, there is no frequency blue-shift of the optical mode for the case of single $P_2$ pulse excitation, even if the single $P_2$ pump fluence exceeds the total fluence of the double pump-pulses (17.5 mJ cm$^{-2}$ in Fig. 3b). This result indicates that the transient state characterized by the double-peak FT spectra is induced only by the coherent action of the optical phonon by virtue of the double pump-pulses.

**Fluence dependence for the SET phase of iPCM.** To check if there is a threshold for the observation of the double-peak FT spectra in the SET phase of iPCM, we present the fluence dependence of a pump–pump–probe experiment for $\Delta t$ = 290 fs for a fixed fluence ratio of $P_1$ and $P_2$ pulses (Fig. 4a). At the lowest fluence of $P_1$ = 5.3 mJ cm$^{-2}$ and $P_2$ = 3.5 mJ cm$^{-2}$ (8.8 mJ cm$^{-2}$ in total), only a single phonon peak at ≈ 3.43 THz was observed (Fig. 4b), which is the same peak position as that observed with only $P_2$ pulse excitation at 10.6 mJ cm$^{-2}$ (see the bottom trace of Fig. 4b). On the contrary, for fluences higher than $P_1$ = 10.6 mJ cm$^{-2}$ and $P_2$ = 6.9 mJ cm$^{-2}$, the double-peak FT spectra emerge accompanied by a blue-shift of the original peak. Thus, the choice of fluences $P_1$ = 10.6 mJ cm$^{-2}$ and $P_2$ = 6.9 mJ cm$^{-2}$ was made because of the lower threshold limit for the observation of the double-peak FT spectral feature in iPCM. As mentioned earlier, irradiation by a single $P_2$ without $P_1$ irradiation to the SET phase cannot induce a transition into the transient double-peak coherent phonon spectra even if its fluence ($P_2$ =18.6 mJ cm$^{-2}$ in Fig. 3d) exceeds the combined fluence of a $P_1$ and $P_2$ pair above the threshold, the total fluence of $P_1$ (10.6 mJ cm$^{-2}$) +$P_2$ (6.9 mJ cm$^{-2}$) = 17.5 mJ cm$^{-2}$ in Fig. 4. This finding demonstrates that the threshold for the appearance of the transient state is substantially decreased by irradiation with double pump-pulses or in all likelihood only double-pulse excitation can provide access to the transient state. It is also noted that the required pump fluence used here (10 – 20 mJ cm$^{-2}$) is significantly lower than those used in the experiments using picosecond or nanosecond laser pulses, in which a fluence of 30–60 mJ cm$^{-2}$ for a 30 ps single pulse[30] or a fluence of ≈ 150 mJ cm$^{-2}$ for a 8 ns single pulse was applied[31].

**Transient reflectivity studies of other phases.** In contrast to the pre-transitional dynamics in the SET phase of iPCM, the phase transformation from the RESET phase of the iPCM was irreversible (Supplementary Note 1 and Supplementary Figs. 2–4). The carrier response suggests that the metastable state after irradiation by the $P_1$ pulse to the RESET phase of iPCM is different from that of the so-called laser-crystallized (LC) structure[26] (Supplementary Note 2 and Supplementary Fig. 4), which should show a much different carrier response from the original phase. Note also that for the case of the poly-crystalline (SET phase) GST alloy, no frequency blue-shift (phonon hardening) was observed in the excited state under double pump-pulse excitation, but only broadening and red-shift of the optical phonon mode was observed on the time scale of a few picoseconds (Supplementary Note 3 and Supplementary Fig. 5). Since photo-excited carriers in semiconductors relax via carrier-phonon scattering (intraband relaxation) and trapping by defects in a few picoseconds[32] (Supplementary Fig. 1), the observed frequency red-shift is interpreted to be the result of phonon softening induced by electronic excitation[26]. From the fact that for the poly-crystalline GST alloy no frequency blue-shift of the optical phonon was observed, it is concluded that the transient state cannot be accessed by the present double-pulse excitation.

**Discussion**
To discuss the pathways of pre-transitional structural dynamics observed in the excited state, we consider the transient structure of iPCM in terms of its bonding nature[7,33]. Given that [Ge$_2$Te$_2$/Sb$_2$Te$_3$]$_{20}$ iPCM sample has the same average composition as Ge$_2$Sb$_2$Te$_5$ (refs. 7 and 14), we refer to a recent theoretical study based on *ab-initio* MD simulation by Li *et al.*, who reported that the phase transformation from the crystalline (SET) to amorphous (RESET) phase in GST alloy occurred over several picoseconds under optical excitation[13]. The authors also claimed that the coordination number of Ge atoms changed from the original six-fold into a mixture of five-fold and four-fold coordination within 450 fs after the removal of 9% of the valence electrons in a GST alloy, followed by the further appearance of three-fold coordination at 3 ps. On the other hand, Simpson *et al.* reported for the iPCM system, the presence of a lower coordination for Ge atoms, namely the possibility of four-fold or three-fold coordination[7]. These studies suggest that in the early stages of the non-thermal pre-phase transformation from the SET to RESET phases under repetitive photo-excitation by double pump-pulses, the bonding coordination around Ge atoms in the iPCM system is coherently modulated and becomes unstable giving rise to two different coordination environments (possibly four-fold and three-fold[7]), providing a plausible explanation as to why a double peak spectral feature develops for the earlier time intervals of $\Delta t$ = 290 to 1,160 fs in Fig. 3b. In contrast to the case of strong photo-excitation by a single pump-pulse, hot-carrier injection on ultrafast time scales of several femtoseconds is suppressed and concurrently similar conditions to a mode-selective vibrational excitation[34] are established under double pump-pulse excitation.

To conclude, we have experimentally explored the photo-excited state of GST phase change materials, iPCM films, by comparison between the cases of single and double pulse excitation using pump-pump-probe femtosecond coherent phonon spectroscopy far from equilibrium. A transient hidden phase characterized by double-peak FT spectra was uncovered for iPCM structures only when a double-pump-pulse sequence was applied, which is interpreted as being due to a mixture of two different Ge coordination environments, which relax within a few picoseconds. Our experimental results also demonstrate that the transition of the SET phase of iPCM to a hidden phase can be achieved with lower total energy by using multiple pulse sequences than for the case of single pulse excitation. These effects will provide a new route for faster and lower threshold phase switching with further testing of the iPCM superlattice structure. Thus, our finding of a non-thermal ultrafast pre-phase transformation in a iPCM film coupled with femtosecond multiple pulse sequences (Supplementary Fig. 6) or multiple terahertz light pulse sequences[35] will provide highly relevant fundamental knowledge for ultrafast optical data processing[36] and for next-generation of ultra-high-speed phase change random access memory (PCRAM).

## Methods

**Fabrication of GST superlattice (iPCM) films.** Recently, Chong *et al.* proposed superlattice-like PCRAM considering the GST system as a composite of the pseudo-binary alloys, namely GeTe and $Sb_2Te_3$ alloys, with individual layers thick enough to maintain the characteristics of each composition[5]. Both faster switching times (< 5 ns) and lower programming currents were found for the superlattice-like PCRAM. More recently, motivated by the need to reduce both the switching speed and energy used, interfacial phase change memory (iPCM) was proposed. Tominaga *et al.* reported the fabrication of a GST superlattice (iPCM) based upon the Ge flip-flop transition mechanism[14]. Using iPCM structures designed from GeTe and $Sb_2Te_3$ layers, a few unit cells thick, they experimentally confirmed very low (only 12% compared to GST alloy films[7]) power operation of phase switching (SET ↔ RESET) in iPCM. The sample used in this paper was a thin film (20 nm-thick) of a prototypical iPCM sample, $[Ge_2Te_2/Sb_2Te_3]_{20}$, which consisted of twenty-repetitive sheet blocks from alternatively deposited 0.5-nm-thick GeTe and $Sb_2Te_3$ layers on a Si-(100) wafer using helicon-wave RF magnetron-sputtering.

**Coherent phonon spectroscopy using a pump-pump-probe sequence.** Coherent phonon spectroscopy (CPS) is a powerful tool to study the ultrafast dynamics of structural phase transitions occurring on ultrafast time scales. It has been applied to a wide variety of materials, such as semimetals and semiconductors[22, 25, 28, 37, 38], and Mott insulators[19]. In CPS, a pump-pulse impulsively generates Raman-active collective atomic vibrations through light-matter coupling. We utilized a 40-fs amplified near-infrared optical pulse (800 nm; 1.55-eV and 100 kHz repetition rate) to excite and monitor coherent lattice vibrations in iPCM films after injection of photo-carriers across the indirect band gap of 0.5 - 0.7 eV (ref. 11). The optical penetration depth at 800 nm was estimated from the absorption coefficient to be ∼ 20 nm, which matches the film thickness. The size of the probe beam after focusing was ∼ 50% smaller than the pump, providing for negligibly small inhomogeneous excitation effects. The maximum photo-generated carrier density was estimated to be $n_{exc} \approx 5.1 \times 10^{21}$ cm$^{-3}$, induced by a single pump-pulse with 10.6 mJ cm$^{-2}$, whose density corresponds to ≈ 2.9% of the total number density of valence electrons ($n_{tot} \approx 1.76 \times 10^{23}$ cm$^{-3}$)[39]. A train of two pump-pulses were generated through a Michelson-type interferometer, in which a motorized stage was installed under the mirrors to adjust the time interval ($\Delta t$) of the temporally separated pump-pulses. The ratio of the $P_1$ and $P_2$ pulses was set to 10:6.5, i) in order to match the phonon amplitude of the initially excited phonon at $\tau$ = 0 fs to that of the second excited phonon at separation times of $\Delta t$ = 290 – 870 fs as was used in the previous study[28]; in the present case, the amplitude of the phonon excited by a $P_1$ pulse at $\tau$ = 0 fs decays to ≈ 65 – 70% in amplitude when $\Delta t$ = 290 – 870 fs, and ii) as it was characteristic of the 2-inch beam splitter with a *p*-polarized incident pump beam. The photo-induced reflectivity change ($\Delta R/R$) was recorded as a function of the time delay ($\tau$) between the pump and probe pulses. The delay was scanned over 10 ps and averaged for 1,000 scans by using an oscillating retroreflector with a 10 Hz scan frequency.

**References**

1. Yamada, N., Ohno, E., Nishiuchi, K. & Akahira, N. Rapid phase transitions of GeTe-Sb$_2$Te$_3$ pseudobinary amorphous thin films for an optical disk memory. *J. Appl. Phys.* **69**, 2849-2856 (1991).

2. Wuttig, M. & Yamada, N. Phase-change materials for rewritable data storage. *Nat. Mater.* **6**, 824-832 (2007).

3. Hegedüs, J. & Elliott, S. R. Microscopic origin of the fast crystallization ability of Ge-Sb-Te phase-change memory materials. *Nat. Mater.* **7**, 399-405 (2008).

4. Ovshinsky, S. R. Reversible electrical switching phenomena in disordered structures. *Phys. Rev. Lett.* **21**, 1450-1453 (1968).

5. Chong, T. C. *et al*. Phase change random access memory cell with superlattice-like structure. *Appl. Phys. Lett.* **88**, 122114 (2006).

6. Shportko, K. *et al*. Resonant bonding in crystalline phase-change materials. *Nat. Mater.* **7**, 653–658 (2008).

7. Simpson, R. E. *et al*. Interfacial phase-change memory. *Nat. Nanotech.* **6**, 501-505 (2011).

8. Kolobov, A. V. *et al*. Understanding the phase-change mechanism of rewritable optical media. *Nat. Mater.* **3**, 703-708 (2004).

9. Caravati, S. *et al*. Coexistence of tetrahedral- and octahedral-like sites in amorphous phase change materials. *Appl. Phys. Lett.* **91**, 171906 (2007).

10. Huang, B. & Robertson, J. Bonding origin of optical contrast in phase-change memory materials. *Phys. Rev. B* **81**, 081204(R) (2010).

11. Lee, B.-S. *et al.* Investigation of the optical and electronic properties of Ge$_2$Sb$_2$Te$_5$ phase change material in its amorphous, cubic, and hexagonal phases. *J. Appl. Phys.* **97**, 093509 (2005).

12. Akola, J. & Jones, R. O. Structural phase transitions on the nanoscale: The crucial pattern in the phase-change materials Ge$_2$Sb$_2$Te$_5$ and GeTe. *Phys. Rev. B*. **76**, 235201 (2007).

13. Li, Xian-Bin. *et al*. Role of electronic excitation in the amorphization of Ge-Sb-Te alloys. *Phys. Rev. Lett.* **107**, 015501 (2011).

14. Tominaga, J. *et al*. What is origin of activation energy in phase-change film? *Jpn. J. Appl. Phys.* **48**, 03A053 (2009).

15. Tominaga, J., Kolobov, A. V., Fons, P., Nakano, T. & Murakami, S. Ferroelectric order control of the Dirac-semimetal phase in GeTe-Sb$_2$Te$_3$ superlattices. *Adv. Mater. Interfaces* **1**, 1300027 (2014).

16. Sundaram, S. K. & Mazur, E. Inducing and probing non-thermal transitions in semiconductors using femtosecond laser pulses. *Nat. Mater.* **1**, 217–224 (2002).

17. Kolobov, A. V. *et al*. Distortion-triggered loss of long-range order in solids with bonding energy hierarchy. *Nat. Chem.* **3**, 311-316 (2011).

18. Polli, D. *et al*. Coherent orbital waves in the photo-induced insulator–metal dynamics of a magnetoresistive manganite. *Nat. Mater.* **6**, 643–647 (2007).

19. Wall, S. *et al*. Ultrafast changes in lattice symmetry probed by coherent phonons. *Nat. Commun.* **3**, 721 (2012).

20. Konishi, M. *et al*. Ultrafast amorphization in $Ge_{10}Sb_2Te_{13}$ thin film induced by single femtosecond laser pulse. *Appl. Opt.* **49**, 3470–3473 (2010).

21. Fons, P. *et al*. Photoassisted amorphization of the phase-change memory alloy $Ge_2Sb_2Te_5$. *Phys. Rev. B* **82**, 041203(R) (2010).

22. Makino, K., Tominaga, J. & Hase, M. Ultrafast optical manipulation of atomic arrangements in chalcogenide glass memory materials. *Opt. Express* **19**, 1260–1270 (2011).

23. Yusupov. R. *et al*. Coherent dynamics of macroscopic electronic order through a symmetry breaking transition. *Nat. Phys.* **6**, 681–684 (2010).

24. Hase, M. *et al*. Forcibly driven coherent soft phonons in GeTe with intense THz-rate pump fields. *Appl. Phys. Lett.* **83**, 4921-4923 (2003).

25. Först, M. *et al*. Phase change in $Ge_2Sb_2Te_5$ films investigated by coherent phonon spectroscopy. *Appl. Phys. Lett.* **77**, 1964-1966 (2000).

26. Hernandez-Rueda, J. *et al*. Coherent optical phonons in different phases of $Ge_2Sb_2Te_5$ upon strong laser excitation. *Appl. Phys. Lett.* **98**, 251906 (2011).

27. Dekorsy, T., Kütt, W., Pfeifer, T. & Kurz, H. Coherent control of LO phonon dynamics in opaque semiconductors by femtosecond laser pulses. *Europhys. Lett.* **23**, 223–228 (1993).

28. Hase, M. *et al*. Optical control of coherent optical phonons in bismuth films. *Appl. Phys. Lett.* **69**, 2474–2476 (1996).

29. Weiner, A. M., *et al*. Femtosecond pulse sequences used for optical manipulation of molecular motion. *Science* **247**, 1317-1319 (1990).

30. Siegel, J., Schropp, A., Solis, J., Afonso, C. N. & Wuttig, M. Rewritable phase-change optical recording in $Ge_2Sb_2Te_5$ films induced by picosecond laser pulses. *Appl. Phys. Lett.* **84**, 2250-2252 (2004).

31. Siegel, J. *et al*. Amorphization dynamics of $Ge_2Sb_2Te_5$ films upon nano- and femtosecond laser pulse irradiation. *J. Appl. Phys.* **103**, 023516 (2008).

32. Zhang, G., Gan, F., Lysenko, S. & Liu, H. Observation of ultrafast carrier dynamics in amorphous $Ge_2Sb_2Te_5$ films induced by femtosecond laser pulses. *J. Appl. Phys.* **101**, 033127 (2007).

33. Kolobov, A. V., Fons, P., Tominaga, J. & Ovshinsky, S. R. Vacancy-mediated three-center four-electron bonds in $GeTe$-$Sb_2Te_3$ phase-change memory alloys. *Phys. Rev. B* **87**, 165206 (2013).

34. Rini, M. *et al*. Control of the electronic phase of a manganite by mode-selective vibrational excitation. *Nature* **449**, 72-74 (2007).

35. Kampfrath, T., Tanaka, K. & Nelson, K. A. Resonant and nonresonant control over matter and light by intense terahertz transients. *Nat. Photon.* **7**, 680–690 (2013).

36. Mihailovic, D. *et al*. Femtosecond data storage, processing, and search using collective excitations of a macroscopic quantum state. *Appl. Phys. Lett.* **80**, 871–873 (2002).

37. Cho, G. C., Kütt, W. & Kurz, H. Subpicosecond time-resolved coherent-phonon oscillations in GaAs. *Phys. Rev. Lett.* **65**, 764–766 (1990).

38. Zeiger, H. J. *et al*. Theory for displacive excitation of coherent phonons. *Phys. Rev. B* **45**, 768–778 (1992).

39. C. Steimer. *et al*. Characteristic ordering in liquid phase-change materials. *Adv. Mater.* **20**, 4535–4540 (2008).

**Acknowledgements**
This work was supported by the X-ray Free Electron Laser Priority Strategy Program (NO. 12013011 and 12013023), from the Ministry of Education, Culture, Sports, Science and Technology of Japan MEXT, and by CREST, Japan Science and Technology Agency (JST).

**Author Contributions**
M. H. and J. T. planned and organized this project. J. T. fabricated the sample. M. H. performed experiments and analyzed the data. M. H., P. F., K. M., A. V. K. and J. T. discussed the results. M. H. and P. F. co-wrote the manuscript.

*Femtosecond structural transformation of phase-change materials far from equilibrium monitored by coherent phonons*

**Figure captions**

**Figure 1 | Coherent phonon spectroscopy in the photo–excited state.** (**a**) Schematic of the pump-pump-probe experiment, showing photo-excitation of the system from the ground state into the excited state with a prepump-pulse ($P_1$ = 10.6 mJ cm$^{-2}$), followed by generation of a coherent phonon in the excited state with another weak pump ($P_2$ = 6.9 mJ cm$^{-2}$), which is monitored by a probe pulse ($P_3$ = 0.2 mJ cm$^{-2}$). (**b**) Schematic of the potential energy surface, together with the local structural change of the iPCM. The ground state of the SET phase (red curve) is characterized by an anharmonic potential, whose activation energy barrier is larger than the difference in the free energy between the SET and RESET (black curve) phases[14]. The photo-excitation ($h\nu$ = 1.55-eV) promotes the ground state into the excited state (blue curve), where the non-thermal phase transformation can be induced and monitored by pump-pump-probe coherent phonon spectroscopy. The local structures in the two ground states were calculated by first principle simulations[22]. The green balls represent Ge atoms, the orange balls are Te atoms, and purple balls are Sb atoms.

**Figure 2 | Time-domain coherent phonon responses in photo-excited iPCM.** (**a**) Transient reflectivity traces observed in the SET phase of an iPCM film for in-phase separation times between $P_1$ (10.6 mJ cm$^{-2}$) and $P_2$ (6.9 mJ cm$^{-2}$) pulses; $\Delta t$ = 290 fs, 870 fs, 1,450 fs, 2,030 fs, and 6,090 fs. The light-blue rectangles represent the coherent phonon signal used for monitoring the excited lattice, which were converted into FT spectra in Fig. 3b. The result for the case without the prepump ($P_1$) is shown at the bottom for reference. (**b**) Transient reflectivity traces observed in the SET phase of iPCM film at various pump fluences from 5.3 to 18.6 mJ cm$^{-2}$ under the single pulse excitation with only the $P_2$ pulse. The coherent phonon oscillations show longer relaxation dynamics at lower fluences, while they show shorter relaxation dynamics at higher fluences.

**Figure 3 | Coherent phonon spectra in photo-excited iPCM.** (**a**) FT spectrum in the SET phase monitored by a weak pump ($P_2$ = 6.9 mJ cm$^{-2}$) and probe ($P_3$ = 0.2 mJ cm$^{-2}$) pulses without the prepump–pulse ($P_1$). The dotted line in (**a**) corresponds to the frequency of the optical mode in the SET phase ($\Omega_{SET}$= 3.48 THz). (**b**) FT spectra obtained from the time-domain data in the excited state at various $\Delta t$ as shown in Fig. 2a. The total fluence applied was $F_{total}$ = 17.5 mJ cm$^{-2}$. The red arrows show the split of the optical mode into doublet peaks at 3.7 THz and 2.55 THz at $\Delta t$ = 290 fs. The dotted lines in (**b**) correspond to the dynamic shift of the 3.7 THz peak down to 3.57 THz at $\Delta t$ = 1,740 fs. The black arrows point out the positions of a peak at 2.55 THz. (**c**) FT spectrum observed at the same spot after the measurement for $\Delta t$ = 6,090 fs in (**b**) monitored without prepump ($P_1$). The dotted line in (**c**) is located at 3.48 THz, indicating the system reverts to the initial SET phase. (**d**) The FT spectra in the SET phase of iPCM obtained from the time-domain data in Fig. 2b. The peak frequency at 5.3 mJ cm$^{-2}$ is 3.52 THz, while at the highest pump fluence of 18.6 mJ cm$^{-2}$ it red-shifts to 3.27 THz as shown by the red arrow over the dotted lines.

**Figure 4 | Pump fluence dependence of pre-transitional dynamics in the SET phase of iPCM.** (**a**) Time-domain signal observed at various total pump fluences for a constant fluence ratio of the $P_1$ and $P_2$ pulses and for a fixed separation time of $\Delta t$ = 290 fs. The bottom trace was obtained by using only the $P_2$ pump pulse before the series of the pump-pump-probe experiment with $P_1$, $P_2$ and $P_3$ pulses, while the top trace was obtained immediately after the measurement of a pump-pump-probe experiment with $P_1$ = 16 mJ cm$^{-2}$ and $P_2$ = 10.6 mJ cm$^{-2}$. (**b**) The corresponding FT spectra obtained from the time-domain data in (**a**). The single peak frequency before the irradiation by the $P_1$ pulse is 3.43 THz. When the fluence was increased beyond $P_1$ = 10.6 mJ cm$^{-2}$ and $P_2$ = 6.9 mJ cm$^{-2}$ (17.5 mJ cm$^{-2}$ in total), a double-peak structure appeared, accompanying by frequency blue-shift to $\approx$ 3.7 THz. The reversible process of these experiments was confirmed by the top trace, showing the peak position is nearly identical to that before exposure to the $P_1$ pulse. The dotted lines represent the position of the peaks at 3.43 and 3.7 THz, respectively.

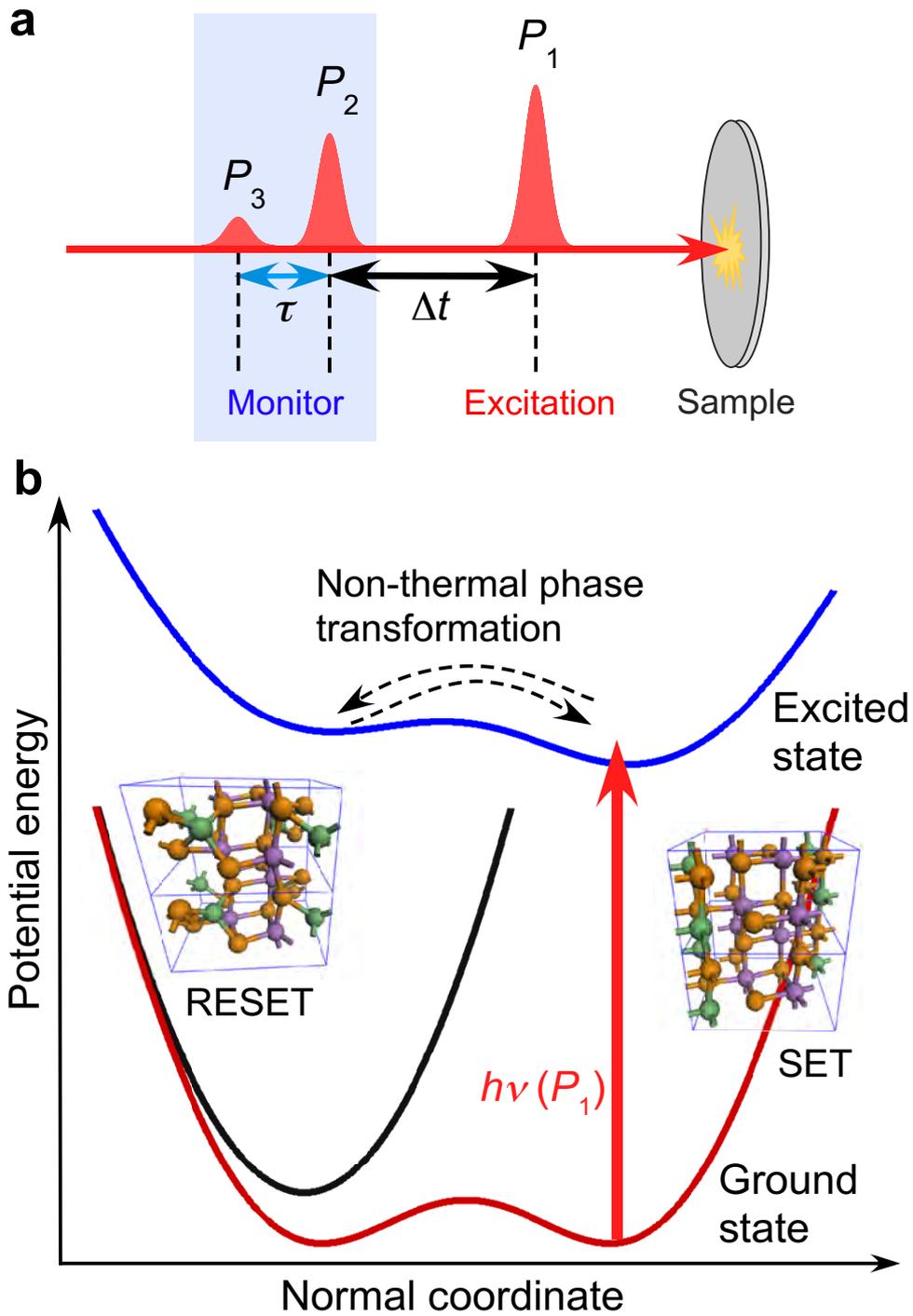

**Fig. 1.** Hase *et al.*

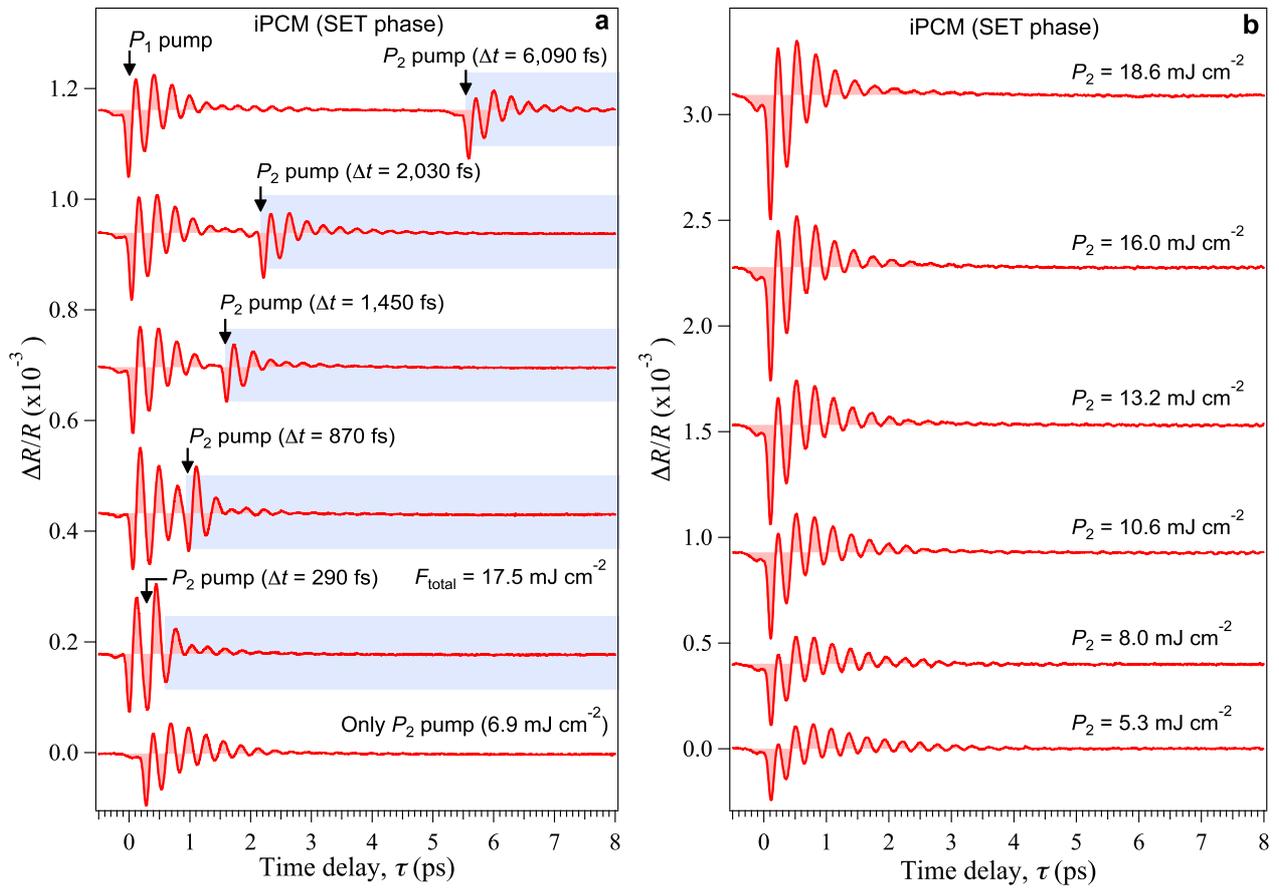

**Fig. 2.** Hase *et al.*

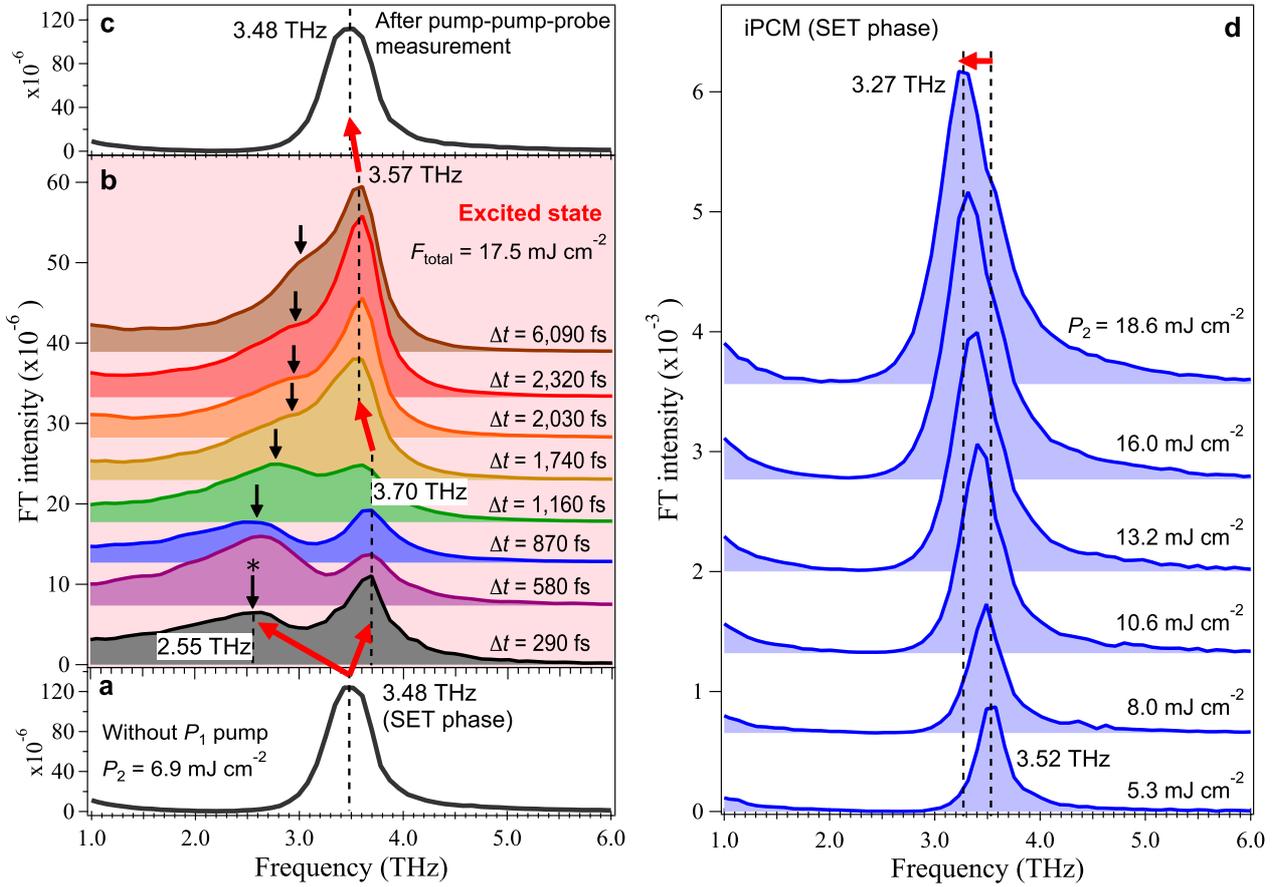

**Fig. 3.** Hase *et al.*

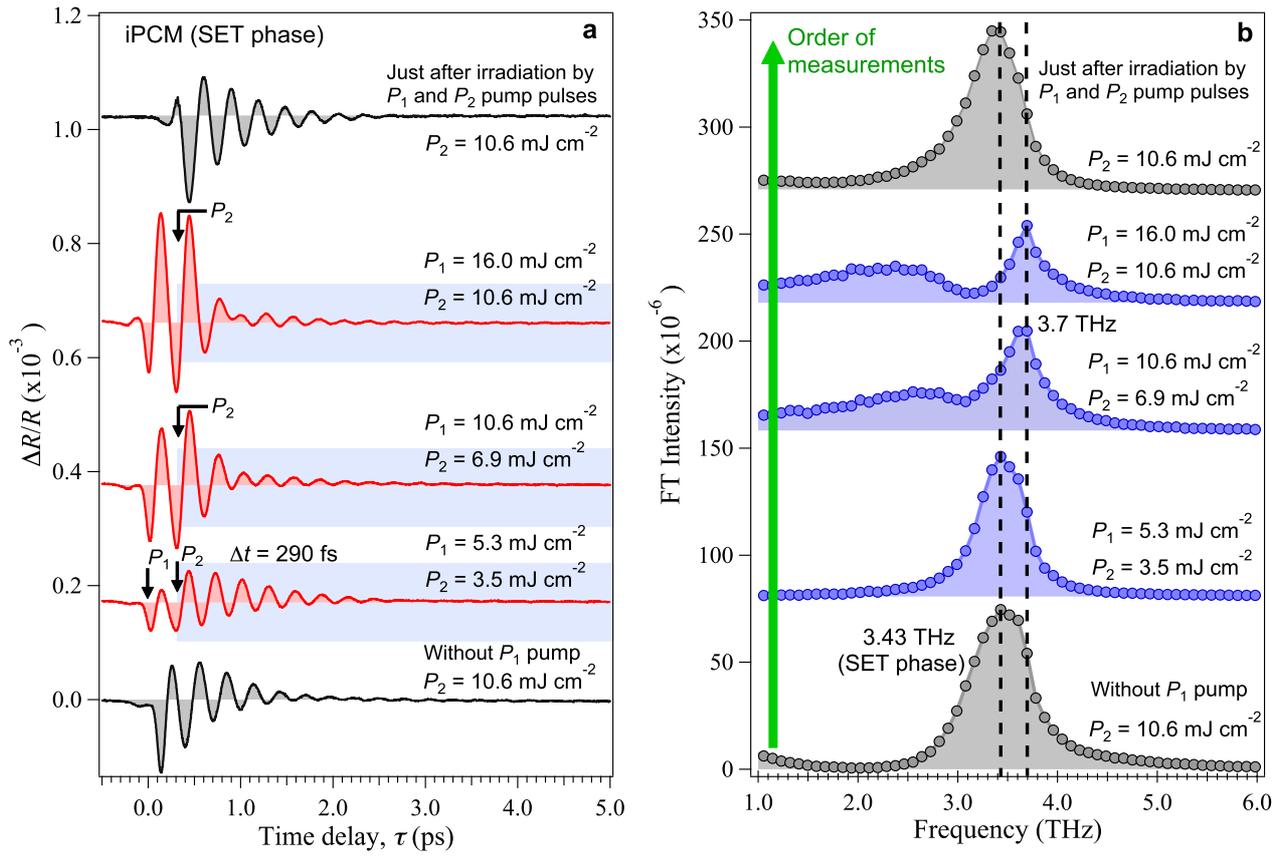

**Fig. 4.** Hase *et al*